# GIBBS POSTERIOR FOR VARIABLE SELECTION IN HIGH-DIMENSIONAL CLASSIFICATION AND DATA MINING[1]

By Wenxin Jiang and Martin A. Tanner

*Northwestern University*

In the popular approach of "Bayesian variable selection" (BVS), one uses prior and posterior distributions to select a subset of candidate variables to enter the model. A completely new direction will be considered here to study BVS with a Gibbs posterior originating in statistical mechanics. The Gibbs posterior is constructed from a risk function of practical interest (such as the classification error) and aims at minimizing a risk function without modeling the data probabilistically. This can improve the performance over the usual Bayesian approach, which depends on a probability model which may be misspecified. Conditions will be provided to achieve good risk performance, even in the presence of high dimensionality, when the number of candidate variables "$K$" can be much larger than the sample size "$n$." In addition, we develop a convenient Markov chain Monte Carlo algorithm to implement BVS with the Gibbs posterior.

**1. Introduction.** The problem of interest here is to predict $y$, a $\{0, 1\}$ response, based on $x$, a vector of predictors of dimension $\dim(x) = K$. We have $D^n = (y^{(i)}, x^{(i)})_1^n$, the observed data with *sample size $n$*, typically assumed to form $n$ i.i.d. (independent and identically distributed) copies of $(y, x)$.

One is often interested in modeling the relation between $y$ and $x$, selecting components of $x$ that are most relevant to $y$, and predicting $y$ using selected information from $x$.

In the approach of *Bayesian variable selection* (*BVS*), one chooses components of $x$ according to some probability distribution (prior and posterior). The BVS approach is very popular for handling high-dimensional data (with

Received February 2007; revised August 2007.

[1]Supported in part by NSF Grant DMS-07-06885.

large dimension $K$, sometimes larger than the sample size $n$), and has had a wide range of successful applications. See, for example, Smith and Kohn (1996), George and McCulloch (1997), Gerlach, Bird and Hall (2002), Lee, Sha, Dougherty, Vannucci and Mallick (2003), Zhou, Liu and Wong (2004) and Dobra, Hans, Jones, Nevins, Yao and West (2004), among others.

For classification purpose, a regression model $p = p(y|x)$ ($y \in \{0, 1\}$) is typically assumed to be logit linear or probit linear and parameterized by a parameter $\beta$, that is, $p(y|x) = \mu^y (1 - \mu)^{1-y}$, where $\mu = \frac{\exp(x^T \beta)}{1 + \exp(x^T \beta)}$ (for logistic regression) or $\int_{-\infty}^{x^T \beta} (2\pi)^{-1/2} e^{-u^2/2} \, du$ (for probit regression). A prior on $p$ is then induced by placing a prior on parameter $\beta$, forcing most of its components to be zero, such that only a low-dimensional subset of $x$ is selected in regression. The corresponding posterior follows a standard Bayesian treatment as $(posterior) \propto (likelihood) \times (prior) \propto \{\prod_{i=1}^n p(y^{(i)}|x^{(i)})\} \times (prior)$. A number of things can be generated from this posterior: parameter $\beta$, conditional density $p(y|x)$, mean function $\mu$, as well as the classification rule (for $y$) $I[\mu > 0.5] = I[x^T \beta > 0]$. Jiang (2007) has shown that under certain regularity conditions, the prior can be specified to render near-optimal posterior performance for density estimation, mean estimation and classification.

The current paper introduces a new direction to BVS. Unlike Jiang (2007), we will construct a modified posterior (called Gibbs posterior) using a risk function of interest (such as the classification error) directly, instead of using the usual likelihood-based Bayesian posterior. We will first focus on the statistical properties (e.g., classification performance) of BVS with a Gibbs posterior. (Section 7 will handle the algorithmic aspects.)

*A problem with the usual Bayesian posterior.* Below, we first demonstrate by a simple example that in case of model misspecification, the usual likelihood-based BVS can provide suboptimal performance. Later our theory will suggest that the proposed BVS with Gibbs posterior can improve over the usual approach, since we will show that the proposed method can still achieve near-optimality in some sense, despite the potential misspecification.

In Jiang (2007), it is assumed that the true model (with density $p^*$) is of a known transformed linear form, say, logit linear, so that $\ln\{p^*(y = 1|x)/p^*(y = 0|x)\}$ is linear in predictors $x_1, \ldots, x_K$, which can be, for example, expressions of $K$ candidate genes.

Suppose we denote the true model by $p^*$ and the set of all logit linear models by $\Lambda$. Then the assumption says $p^* \in \Lambda$. What if this assumption (e.g., logit linearity) is not true, so that higher-order terms and interactions are important but not included? That is, what if the prior proposes densities in $\Lambda$, but the true density $p^* \notin \Lambda$? Then the usual likelihood-based posterior will propose densities that are consistent for (often close to) $p_{KL} = \arg\min_{p \in \Omega} \int p^* \ln(p^*/p)$, a minimizer of KL (Kullback–Leibler) difference $KL(p^*, p) = \int p^* \ln(p^*/p)$, under some regularity conditions—see Kleijn and van der Vaart ([2006](#)). However, this limit $p_{KL}$ (of the usual Bayesian inference) may have a suboptimal risk performance! That is, one can have $\tilde{R}(p_{KL}) > \inf \tilde{R}_{p \in \Omega}(p)$ for a risk function of practical interest such as the classification error $\tilde{R}(p) = P^*\{y \neq I[p(y=1|x) > 0.5]\}$. [See Devroye et al. ([1996](#)), Section 4.6 (least squares) and Section 15.2 (maximum likelihood) for some related comments.]

*An example.* Consider the case when $P^*(x = \pm 1) = \lambda$, $P^*(x = 0) = 1 - 2\lambda$ for some $\lambda \in (0, 0.25)$, $P^*(y=1|x) = 1 - P^*(y=0|x) = I[x \neq 0]$, which define the true density $p^*$. Let $\Lambda$ be the set of densities from logistic regression with $p(y=1|x) = e^{\alpha + x\beta}/(1 + e^{\alpha + x\beta})$, $\alpha, \beta \in \Re$. Note that $p^* \notin \Lambda$. The logistic regression model is misspecified.

According to the KL criterion, the best choice $p_{KL}(y=1|x) = 2\lambda < 0.5$. This is what the usual posterior-based logistic regression will converge to according to Kleijn and van der Vaart ([2006](#)). The resulting classifier $C_{KL}(x) = I[p_{KL}(y=1|x) > 0.5] = 0$ always predicts 0 and the resulting classification error $\tilde{R}(p_{KL}) = 2\lambda$. On the other hand $\tilde{R}(p)$ is minimized to be $\tilde{R}(p_R) = \lambda$ at $p = p_R$, where, for example, $p_R(y=1|x) = e^{x-0.7}/(1 + e^{x-0.7})$, which corresponds to a linear classification rule $C_R = I[p_R(y=1|x) > 0.5] = I[x - 0.7 > 0]$.

For example, when $\lambda = 0.125$, $\tilde{R}(p_{KL}) = 0.25 > \tilde{R}(p_R) = 0.125$, even though both $p_{KL}, p_R \in \Lambda$. So, when the model is misspecified, the usual posterior-based logistic regression is not reliable; it produces suboptimal classification error even from among the misspecified logistic regression models $\Lambda$.

In such situations with model misspecification, a modified posterior directly related to the risk function of interest, called the Gibbs posterior, can still perform very well, unlike the usual likelihood-based (Bayesian) posterior. In Section [2](#), we discuss the Gibbs posterior for risk minimization. What is the Gibbs posterior? How is it interpreted? In addition, we incorporate a smoothed risk function to the Gibbs posterior for computing ease. Then we describe how to evaluate the risk performance of the proposed method in two scenarios in Section [3](#). In Section [4](#), we introduce for the first time the framework of BVS with the Gibbs posterior, which is intended to effectively handle high-dimensional data. Then we provide some results on classification performance in Section [5](#), which show that BVS with the Gibbs posterior can perform very well in some sense, despite high dimensionality, without assuming that the true model is logit linear. These results use a special kind of normal-binary prior. The results are proved under a more general framework in Section [6](#), using more general conditions on the prior and on the risk function. In particular, this covers more general risk functions used in data mining (in addition to classification performance). Some preparatory results for the proofs will be presented in Section 6.1.

Section 7 will handle the algorithmic aspects of sampling from the Gibbs posterior with variable selection. We will show that a convenient and modular Markov chain Monte Carlo (MCMC) algorithm is available based on data augmentation [Tanner and Wong (1987)], so that all sampling steps are based on standard distributions.

**2. Risk minimization with Gibbs posterior.** The previous example shows that for the purpose of minimizing the classification error $\tilde{R}(p)$ over the logit linear models $p \in \Lambda$, it is preferable not to use $p$ proposed from the usual likelihood-based (Bayesian) posterior over $p \in \Lambda$ of the form $(posterior) \propto (likelihood) \times (prior)$.

Note that for logit linear models $p \in \Lambda$, the classification rule $I[p(y = 1|x) = 0.5] = I[x^T\beta > 0]$ forms a linear decision rule (indexed by $\beta$). We are interested in minimizing $\tilde{R}(p) = P^*\{y \neq I[p(y = 1|x) > 0.5]\} = P^*\{y \neq I(x^T\beta > 0)\} \equiv R(\beta)$. For this purpose, there is really no need to assume a probability model $p$ and interpret $\beta$ as a parameter associated with $p$. Instead, we can think of $\beta$ as indexing a linear decision rule $I[x^T\beta > 0]$ and try to minimize a risk function $R(\beta) = P^*\{y \neq I(x^T\beta > 0)\}$.

For this purpose, it is better to use a *Gibbs posterior* over $\beta \in \Omega$ for some parameter space $\Omega \subset \Re^{K_n}$:

$$\omega(d\beta|D^n) = w(d\beta|D^n)\pi(d\beta) = e^{-n\psi R_n(\beta)}\pi(d\beta)\Big/ \int_{\beta \in \Omega} e^{-n\psi R_n(\beta)}\pi(d\beta),$$

where $\pi$ is a prior over $\beta \in \Omega$, and $\psi > 0$ is a constant to be explained later in this section.

Here $R_n$ is a sample version of $R$ depending on (i.i.d.) data $D^n$. Examples include:

(i) $R_n = n^{-1}\sum_{i=1}^n I[y^{(i)} \neq A_i] = -\psi^{-1}n^{-1}\sum_{i=1}^n \log\{A_i e^{\psi(y^{(i)}-1)} + (1 - A_i)e^{-\psi y^{(i)}}\}$, where $A_i = I[p(y^{(i)} = 1|x^{(i)}) > 0.5] = I[(x^{(i)})^T\beta > 0]$;

(ii) $R_n = -\psi^{-1}n^{-1}\sum_{i=1}^n \log\{\Phi_i e^{\psi(y^{(i)}-1)} + (1 - \Phi_i)e^{-\psi y^{(i)}}\}$, where $\Phi_i = \Phi(\sigma_n^{-1}(x^{(i)})^T\beta)$, $\Phi$ is the standard normal cumulative density function and $\sigma_n$ is a scaling factor.

Choices (ii) and (i) are close when $\sigma_n \to 0$ but choice (ii) makes $R_n$ smooth in $\beta$! Later on (in Remark 2 and Section 7) we will see that $R_n$ in (ii) is related to a mixture model and can be used to simplify the posterior simulation.

The Gibbs posterior density $w$ (with respect to the prior $\pi$) minimizes a combination of an averaged sample risk and a penalty against the "change in knowledge" (from prior $\pi$ to posterior $w\pi$). Such an interpretation is given in Zhang (2006a), Proposition 5.1 and Zhang (2006b), Section IV.

PROPOSITION 1 [Zhang (2006a, 2006b)].    *The Gibbs posterior density*

$$w = e^{-n\psi R_n} \Big/ \int_{\beta \in \Omega} e^{-n\psi R_n} \pi(d\beta)$$

*minimizes* $\int_{\beta \in \Omega} w n R_n(\beta) \pi(d\beta) + \psi^{-1} KL(w\pi(d\beta), \pi(d\beta))$ *over all densities* $w$ *on* $\Omega$ *with respect to the prior* $\pi$. *Here* $KL(w\pi(d\beta), \pi(d\beta)) = \int_{\beta \in \Omega} w(\log w) \times \pi(d\beta)$.

The parameter $\psi^{-1}$ in the Gibbs posterior is related to the temperature in statistical mechanics and was used, for example, in Geman and Geman (1984) when studying simulated annealing. The case of zero or very low temperature corresponds to deterministic empirical risk minimization. Allowing nonzero temperatures results in a more general setup of random estimation and allows potential improvement over the deterministic approach. The temperature $\psi^{-1}$ is typically treated as a given constant [e.g., in Zhang (2006b)], but when necessary, an optimal temperature [e.g., Zhang (1999)] may be obtained by, for example, cross validation, as mentioned in Zhang (2006b).

This framework of the Gibbs posterior has been overlooked by most statisticians for a long time, especially when compared to the long-term popularity of the (likelihood-based) Bayesian posterior. Recently, however, the sequence of papers by Zhang (1999, 2006a, 2006b) have laid a foundation for understanding the statistical behavior of the Gibbs posterior, which we believe will open a productive new line of research. While Zhang's (2006b) work concerns fundamental convergence properties of the Gibbs posterior in general, our work focuses on the aspect of variable selection, which is important for handling high-dimensional data with the Gibbs posterior (see the counterexample in Section 4.1). In addition, we allow a computation-friendly smoothed risk function $R_n$ to be used in a proposed algorithm later. Also, Zhang (2006b) has considered the case with high temperature (small $\psi$), while our result holds for any $\psi$, even for low temperature, which might be of interest. It might be of interest to use, for example, a low temperature to recover the results from empirical risk minimization (or maximize the Gibbs posterior) using an approach similar to simulated annealing. Also, we expect that the MCMC algorithm in Section 7 may have better convergence behavior in the low-temperature case since it will depend on the data more heavily.

**3. Critical questions on risk performance: two scenarios.**  Define $P_{\beta, D}$ as the joint distribution based on $p^*(D^n) w(\beta | D^n)$, with $E_{\beta, D}$ being the corresponding expectation. This corresponds to randomly generating data $D^n$ from the true density $p^*$ and then selecting $\beta$ randomly from the Gibbs posterior $\omega(d\beta|D^n)$. The word "often" in the following statements refers to a high probability in $P_{\beta,D}$.

Let $R(\beta)$ be a risk function such as $R(\beta) = P^*[y \neq I(x^T\beta > 0)]$. We will denote $\inf_{\beta \in B} R(\beta) = \inf R(B)$ for a set of decision rules $I(x^T\beta > 0)$ indexed by $\beta \in B$. We will address the following question.

*With high-dimensional data $[K = \dim(x) \gg n]$, will the Gibbs posterior (with variable selection) often lead to a good risk performance which is competitive to all models in $B$? That is, will the method often propose $\beta$ such that $R(\beta) \leq \inf_{\beta \in B} R(\beta) + (small \ \delta)$?*

We will answer this question in two scenarios with a trade-off between the strengths of assumptions and results. Scenario I will involve more assumptions (including a sparseness assumption) but better risk performance (competitive to a bigger set of models $B$). Scenario II will involve fewer assumptions (allowing nonsparse cases) but will guarantee a less optimal risk performance (competitive to a smaller set of models $B$).

The Scenario-I treatment uses a bigger set $B = \Omega$, which here corresponds to the set of *all* linear decision rules (see Section 4.2 for a more precise definition). We will try to show posterior performance competitive to *all* linear rules ["often" $R(\beta) \leq \inf_{\beta \in \Omega} R(\beta) + \delta$]. We will typically need to assume that a best linear rule in $\Omega$ satisfies some sparseness conditions: $\beta_R \in H$, where $\beta_R$ is a minimizer of $R$ over $\Omega$ and $H$ is a "sparse subset" of $\Omega$ satisfying some sparseness conditions.

The Scenario-II treatment will address a smaller set $B = H$, which corresponds to some set of *sparse* linear decision rules. We will try to show posterior performance competitive to all *sparse* linear rules ["often" $R(\beta) \leq \inf_{\beta \in H} R(\beta) + \delta$]. Although the results are competitive to fewer rules, the assumptions needed are also less restrictive: we no longer need to assume that a best linear rule is sparse ($\beta_R \in H$).

This study is about a "nearly best" performance over a set of decision rules in $B$, while not assuming a true probability model for data. This is similar to the "persistence" study for risk minimization by Greenshtein (2006), in a frequentist approach. We now are considering the Bayesian analog so the use of the prior $\pi$ will also matter, which will form part of the regularity conditions.

The questions raised in this section will be answered in the next two sections.

## 4. BVS with a Gibbs posterior.
To answer the questions in Section 3 on risk performance, we first give an example to show the need of variable selection in the high-dimensional case. Without variable selection, even if the Gibbs posterior is used, the risk performance may still be very poor when $K = \dim(x) \gg n$. With variable selection (to be described in Sections 4.2 and 4.3), however, we will show later (in Section 5) that the risk performance can be very good in the two scenarios described in Section 3.

4.1. *An example: high-dimensional classification with Gibbs posterior without variable selection.* Suppose the true model $P^*$ is specified by $y = I[z = 1]$ where $z$ is uniform over $\{1/K, \ldots, K/K\}$. Define $x$ as the vector with components $x_j = I[z = (K+1-j)/K]$, $j = 1, \ldots, K$. Note that the best linear classification rule can be written as $I[x^T\beta > 0]$ where $\beta = (1, 0, \ldots, 0)^T$. This classification rule can be equal to $I[z = 1] = y$ and therefore has classification error $R(\beta) = P^*[y \neq I(x^T\beta > 0)] = P^*[y \neq y] = 0$. (Note that $x^T\beta = \beta_{K+1-Kz}$.) Such a perfect performance can be approximately achieved due to the results later using variable selection with the Gibbs posterior. (See, e.g., Section 5.) However, without variable selection, the use of the Gibbs posterior alone will not guarantee a good classification error.

For example, suppose according to the prior $\pi$, $\beta_j$'s are i.i.d. $N(0,1)$ (or more generally, any independent symmetric distributions which have $\pi[\beta_j > 0] = \pi[\beta_j \leq 0]$). Suppose the Gibbs posterior $\propto e^{-n\psi R_n} \times \pi$ where $R_n$ depends on $\beta$ through $x^{(i)T}\beta$ ($= \beta_{K+1-Kz^{(i)}}$), $i = 1, \ldots, n$, where $(x^{(i)})_j = I[z^{(i)} = (K+1-j)/K]$, $j = 1, \ldots, K$, and $(y^{(i)}, z^{(i)})_1^n$ (data) and $(y, z)$ form an i.i.d. sample. Note that the posterior for $\beta_j$ will only be updated by data if $j \in \Delta \equiv \{K+1-Kz^{(i)}\}_{i=1}^n$.

Consider the expected classification error $EP^*_{y,x}[y \neq I(x^T\beta > 0)] = EP^*_{y,z}[y \neq I(\beta_{K+1-Kz} > 0)]$ (where $E = E^*_{(y^{(i)}, z^{(i)})_1^n} E_{\beta|(y^{(i)}, z^{(i)})_1^n}$). This is the "overall" probability of misclassification $\tilde{P}[y \neq I(x^T\beta > 0)] = \tilde{P}[y \neq I(\beta_{K+1-Kz} > 0)]$, where $\beta$ is also random, in addition to the random $y$ and $z$'s. Here, the distribution $\tilde{P}$ is specified by noting that $(y^{(i)}, z^{(i)})_1^n$ are i.i.d. from the true model $P^*$, $\beta|(y^{(i)}, z^{(i)})_1^n$ follows the Gibbs posterior, and $(y, z)$ denotes an independent future observation from $P^*$.

Suppose $z \notin \{z^{(i)}\}_1^n$; then the posterior for $\beta_{K+1-Kz}$ will not be updated by data $(y^{(i)}, z^{(i)})_1^n$. So assuming the event $z \notin \{z^{(i)}\}_1^n$, the conditional probability $\tilde{P}[y \neq I(\beta_{K+1-Kz} > 0)|z, \{z^{(i)}\}_1^n]$ is 0.5, since it is determined by the (un-updated) prior of $\beta_{K+1-Kz}$ which is symmetric about 0. Therefore the probability $\tilde{P}\{[y \neq I(x^T\beta > 0)] \cap [z \notin \{z^{(i)}\}_1^n]\}$ is $0.5P^*[z \notin \{z^{(i)}\}_1^n] \geq 0.5(1 - n/K)$, which can be close to 0.5 for $K \gg n$. This also forms a lower bound of $\tilde{P}[y \neq I(x^T\beta > 0)]$, which is bounded below by $\tilde{P}\{[y \neq I(x^T\beta > 0)] \cap [z \notin \{z^{(i)}\}_1^n]\}$.

Therefore, without variable selection, the expected classification error can be close to 50% when $K \gg n$, even if the Gibbs posterior is used.

We now consider applying BVS with Gibbs posterior for classification, when subsets of candidate variables are used to effectively handle high-dimensional data.

4.2. *A parameterization.* Consider a decision rule $I(x^T\beta > 0)$ for $\beta \in \Re^{K_n}$ ($x$ can include the constant component 1). The risk can be, for example, the misclassification probability $R(\beta) = P^*\{y \neq I(x^T\beta > 0)\}$. It is noted that the decision rule $I(x^T \beta > 0)$ and the risk $R(\beta)$ are not changed under the rescaling of $\beta$. Following the approach of Horowitz (1992), we suppose it is possible to use a standardization with $|\beta_1| = 1$ or $\beta_1 \in \{\pm 1\}$, and define $\beta^T = (\beta_1, \tilde{\beta}^T) \in \{\pm 1\} \times \Re^{K_n - 1}$, and correspondingly $x^T = (x_1, \tilde{x}^T)$.

Let $\Omega_n$ denote the (standardized) parameter space $\Omega_n = \{\pm 1\} \times \Re^{K_n - 1}$. Characterize $\beta$ by $(\gamma, \beta_\gamma)$ where $\gamma = (\gamma_j)_1^{K_n}$ is the "model" indicator with $\gamma_j = I[\beta_j \neq 0]$ ($\gamma_1 = 1$), telling which components of $\beta$ are nonzero. For any vector $v$, the notation $v_\gamma$ denotes the subset of $v_j$'s with $\gamma_j = 1$.

Note that in this parameterization, $x_1$ is always contained in the decision rule with coefficient being $\pm 1$. It can be a variable that we always want to keep for decision-making due to some practical considerations. We can still allow $x_1$ to have effectively very small impact on classification, by allowing other $\tilde{\beta}$ coefficients to be much larger. Adopting such a standardization reduces the redundancy of parameterization and can improve the convergence of the algorithms when simulating the Gibbs posterior.

The Gibbs posterior is induced by a prior $\pi$ on $\beta \in \Omega$, which could be equivalently specified by putting a prior on the parameters $(\gamma, \beta_\gamma)$. Then a Gibbs posterior is obtained as $\omega(d\beta | D^n) \propto e^{-n\psi R_n(\beta)} \pi(d\beta)$ as described in Section 2. Below we will first consider a normal-binary prior for $(\gamma, \beta_\gamma)$.

4.3. *A prior specification (normal-binary).*  For a vector $v = (v_j)_1^d$, we will denote its $\ell_p$ norm ($p = 1, 2, \ldots$) as $|v|_p = (\sum_{j=1}^d v_j^p)^{1/p}$, its $\ell_\infty$ norm as $|v|_\infty = \sup_{j=1}^d |v_j|$, and its $\ell_0$ norm as $|v|_0 = \sum_{j=1}^d I[|v_j| > 0]$.

Suppose $\beta \in \Omega_n$, with standardization $|\beta_1| = 1$ as described above. Suppose for the prior $\pi$, $(\gamma_j)_{j=2}^{K_n}$ (the "model" indicators) are i.i.d. binary with selection probability $\lambda_n$ and size restriction $\bar{r}_n$. Conceptually, one first generates $\breve{\gamma} = \breve{\gamma}_1^{K_n}$ where $\breve{\gamma}_1 = 1$, and $\breve{\gamma}_2^{K_n}$ are i.i.d. binary with selection probability $\lambda_n$. Then set $\gamma = \breve{\gamma}$ only when $|\breve{\gamma}|_1 \leq \bar{r}_n$. Suppose conditional on $\gamma$, $\beta_1$ is independent of $\tilde{\beta}_\gamma$ [the subset of $(\beta_j)_2^{K_n}$ with $\gamma_j = 1$], $\beta_1 | \gamma = \pm 1$ with probability 0.5 each, and $\tilde{\beta}_\gamma | \gamma \sim N(0, V_\gamma)$, according to the prior $\pi$.

## 5. Results on risk performance for BVS with Gibbs posterior.

This section will address the risk performance in the two scenarios described in Section 3, when BVS is applied to the Gibbs posterior as described in Sections 4.2 and 4.3. The risk function $R(\beta)$ here is the classification error, while the Gibbs posterior is constructed from the smooth sample risk $R_n(\beta)$ as described in Section 2 [choice (ii)].

Define the following collection of conditions. Different conditions will be used from this collection for different results, to enable a compressed description of many results.

0′. The candidate variable $x_j$'s are standardized to be between $\pm 1$ for all $j$.

$0''$. The conditional density $p(x_1|\tilde{x})$ with respect to the Lebesgue measure exists for all $x$ and is bounded above by a constant $S > 0$.

$1'$. The rate $\delta_n$ is smaller than 1 and larger than $n^{-1/2}\log n$ in order. ($1 \succ \delta_n \succ n^{-1/2}\log n$.)

$3'$. The dimension $K_n = \dim(x)$ is high and is polynomial in $n$. ($n \prec K_n \prec n^\alpha$ for some $\alpha > 1$.)

$(\sigma)$. The smoothing parameter $\sigma_n$ used in a sample version of $R_n$ decreases to zero in some way as $n$ increases. $[(n/\log n)^{1/2} \overset{\prec}{\sim} \sigma_n^{-1} \prec n^{q''}$ for some $q'' > 1/2$.]

$(V)$. The eigenvalues of prior variance $V_\gamma$ and its inverse are bounded as "model" size $|\gamma|_1$ grows. $[\max\{ch_1(V_\gamma), ch_1(V_\gamma^{-1})\} \leq B$ for some constant $B > 0$, for all large $|\gamma|_1$.]

$(r_\delta)$. The prior size restriction (denoted as $\bar{r}_n$ in Section 4.3) and the prior expectation of "model" size (before size restriction, which is about $\lambda_n K_n$) grow with $n$ in some slow ways: $M' n\delta_n^2/(\log n)^2 \leq \lambda_n K_n \leq \bar{r}_n = \lceil Mn\delta_n^2/(\log n)^2\rceil$ for some $M > 1$ and $M' > 0$. (Here $\lceil\cdot\rceil$ denotes the integer part.)

Finally, we define a collection of "sparse subset" $H$'s of the linear decision rules $\Omega$, which will be used in a condensed statement of many different results.

Let $H_b$ be a "sparse set of rules" of at most $n\delta_n^2/(\log n)^2$ variables with coefficients at most $C$ (some constant): $H_b = \{\beta \in \Omega_n : \sum_j I[|\tilde{\beta}_j| \neq 0] \leq n\delta_n^2/(\log n)^2, \sup_j |\tilde{\beta}_j| \leq C\}$.

Let $H_m$ and $H_E$ be sparse sets satisfying some $\ell_1$ summability conditions with various types of tail behavior (polynomial with power $m$ and exponential, resp.) The formal definitions are:

$H_m = \{\beta \in \Omega_n : \sum_{j \leq K_n} |\tilde{\beta}_{(j)}| \leq C, \sum_{j > r} |\tilde{\beta}_{(j)}| \leq r^{-m}$ for all $r \geq q\}$ for some constants $m, q, C > 0$;

$H_E = \{\beta \in \Omega_n : \sum_{j \leq K_n} |\tilde{\beta}_{(j)}| \leq C, \sum_{j > r} |\tilde{\beta}_{(j)}| \leq e^{-C''r}$ for all $r \geq q\}$ for some constants $q, C, C'' > 0$.

(We use $\tilde{\beta}_{(j)}$ to denote the component of $\tilde{\beta}$ that has the $j$th largest absolute value.)

Let $H_{1,2,3} \supset H_b$ be three other sparse sets, which have at most about $n\delta_n^2/(\log n)^2$ possibly large $\beta$-coefficients, while allowing many more other $\beta$-coefficients to be small and nonzero. The mathematical details are given below:

$H_1 = \{\beta \in \Omega_n : \sum_{j \leq n\delta_n^2/(\log n)^2} |\tilde{\beta}_{(j)}|^2 \leq C^2 n\delta_n^2/(\log n), \sum_{j > n\delta_n^2/(\log n)^2} |\tilde{\beta}_{(j)}| \leq C'\delta_n/(\log n)\}$;

$H_2 = \{\beta \in \Omega_n : \sup_{j \leq n\delta_n^2/(\log n)^2} |\tilde{\beta}_{(j)}| \leq C\sqrt{\log n}, \sum_{j > n\delta_n^2/(\log n)^2} |\tilde{\beta}_{(j)}| \leq C'\delta_n/(\log n)\}$;

$H_3 = \{\beta \in \Omega_n : \sum_{j \leq K_n} |\tilde{\beta}_{(j)}| \leq C, \sum_{j > n\delta_n^2/(\log n)^2} |\tilde{\beta}_{(j)}| \leq C'\delta_n/(\log n)\}$ for some constants $C, C' > 0$.

The following proposition addresses the risk performance of BVS (with a Gibbs posterior) in two scenarios described in Section 3. The results concern the use of the Gibbs posterior $\omega(d\beta|D^n)$ based on $R_n$, under the probability distribution $P_{\beta,D}$ [based on $p^*(D^n)\omega(d\beta|D^n)$] and the corresponding expectation $E_{\beta,D}$.

PROPOSITION 2 (Risk performance). (i) *(Scenario* II*; "exponentially sparse" $H_E$.) Assuming conditions* $0'$, $0''$, $3'$, $(\sigma)$, $(V)$ *and* $(r_\delta)$, *where* $\delta_n = n^{-1/2}(\log n)^2$, *we have*
$$R - \inf R(H_E) \leq cn^{-1/2+\xi} \text{ with } P_{\beta,D}\text{-probability tending to 1 as } n \to \infty,$$
*and* $E_{\beta,D}R - \inf R(H_E) \leq cn^{-1/2+\xi}$ *for all large enough* $n$, *for any* $\xi > 0$, *for some* $c > 0$.

(ii) *(Scenario* I*; "exponentially sparse" $H_E$.) Suppose in addition that* $\inf_{\beta\in\Omega_n} R(\beta)$ *is reached at some* $\beta_R \in H_E$ *(a best rule in* $\Omega_n$ *satisfies the sparsity condition in $H_E$). Then* $R - \inf R(\Omega_n) \leq cn^{-1/2+\xi}$ *with* $P_{\beta,D}$*-probability tending to 1 as* $n \to \infty$, *and* $E_{\beta,D}R - \inf R(\Omega_n) \leq cn^{-1/2+\xi}$ *for all large enough* $n$, *for any* $\xi > 0$, *for some* $c > 0$.

(i)$'$ *(Scenario* II*; "polynomially sparse" $H_m$.) Assuming conditions* $0'$, $0''$, $3'$, $(\sigma)$, $(V)$ *and* $(r_\delta)$, *where* $\delta_n = n^{-m/(2m+1)}(\log n)^2$, *we have* $R - \inf R(H_m) \leq cn^{-m/(2m+1)+\xi}$ *with* $P_{\beta,D}$*-probability tending to 1 as* $n \to \infty$, *and* $E_{\beta,D}R - \inf R(H_m) \leq cn^{-m/(2m+1)+\xi}$ *for all large enough* $n$, *for any* $\xi > 0$, *for some* $c > 0$.

(ii)$'$ *(Scenario* I*; "polynomially sparse" $H_m$.) Suppose in addition that* $\inf_{\beta\in\Omega_n} R(\beta)$ *is reached at some* $\beta_R \in H_m$ *(a best rule in* $\Omega_n$ *satisfies the sparsity condition in $H_m$). Then* $R - \inf R(\Omega_n) \leq cn^{-m/(2m+1)+\xi}$ *with* $P_{\beta,D}$*-probability tending to 1 as* $n \to \infty$, *and* $E_{\beta,D}R - \inf R(\Omega_n) \leq cn^{-m/(2m+1)+\xi}$ *for all large enough* $n$, *for any* $\xi > 0$, *for some* $c > 0$.

Therefore (i) suggests that the Gibbs posterior will lead to performance in $R$ that is no worse than the best performance among the sparse linear rules in $H_E$, up to a rate close to $n^{-1/2}$, despite the high dimension $K_n$ which can be, for example, $n^{10}$. Result (ii) says that if a best linear rule is sparse in $H_E$, then the performance actually is no worse than the best linear rules in $\Omega_n$, up to the same rate despite the high dimension.

When the sparsity conditions from $H_E$ are relaxed to $H_m$, the rate becomes about $n^{-m/(2m+1)}$, which is still not deteriorating as $\dim(x) = K$ increases (even when $K \gg n$). This is in contrast to some other situations (such as regression without variable selection, or piecewise constant models) which have rates deteriorating as the dimension $K$ increases.

The above proposition involves sparse rules that require a bounded $\ell_1$-sum of the $\beta$-coefficients. This limits the number of "potentially important" (or "possibly large") coefficients to be bounded (in $n$). The next proposition generalizes this and allows some other sparse rules, where the number of "possibly large" coefficients can grow in $n$ in some way that affects the convergence rate.

PROPOSITION 3 (Risk performance; other sparse cases). (i) *(Scenario* II; *other sparse cases.) Under conditions 0′, 0″, 3′, (σ), (V), ($r_δ$), with $δ_n$ satisfying 1′, we have $R - \inf R(H_{1,2,3,b}) \le cδ_n$ with $P_{β,D}$-probability tending to 1 as $n \to \infty$, and $E_{β,D}R - \inf R(H_{1,2,3,b}) \le cδ_n$ for all large enough $n$, for some $c > 0$.*

(ii) *(Scenario* I; *other sparse cases.) If in addition $\inf_{β \in Ω_n} R(β)$ is reached at some $β_R \in H_{1,2,3,b}$ (a best model in $Ω_n$ satisfies the sparsity condition in $H_{1,2,3,b}$, resp.), then $R - \inf R(Ω_n) \le cδ_n$ with $P_{β,D}$-probability tending to 1 as $n \to \infty$, and $E_{β,D}R - \inf R(Ω_n) \le cδ_n$ for all large enough $n$, for some $c > 0$.*

Note that there is some compromise between the convergence rate $δ_n$ and the number $v_n = nδ_n^2/(\log n)^2$ (the integer part of) which is the number of "possibly large" $\bar{β}$-coefficients allowed in the "sparse set" $H_{1,2,3,b}$. When $δ_n$ is "precise" or small (such as about $n^{-0.49}$), then $v_n$ is small (about $n^{0.02}$). When $δ_n$ is "rough" or large (such as $n^{-0.01}$), $v_n$ is large (about $n^{0.98}$).

Propositions 2 and 3 will be proved in a more general context of data mining (which need not be classification), as Proposition 5 in Section 6 below, where we also accommodate more general priors (which need not be normal-binary).

**6. Proofs and results for more general priors and risk functions.** Some of the proofs below utilize some preparatory results that will be presented in Section 6.1.

Define the following conditions and notation. Different subsets of these conditions will be used later when formulating different results.

(RnR): For $h \in (0,1)$ and $q > 0$, denote $p_0 = (e^{ψhq} - 1)/(e^{ψq} - 1)$, $p_1 = (1 - e^{-ψhq})/(1 - e^{-ψq})$, $\Phi_i = \Phi(x^{(i)T}β/σ_n)$, $A = I(x^Tβ > 0)$. Let $R_n = -(nψ)^{-1} \sum_{i=1}^n \ln\{\Phi_i p_1^{y^{(i)}}(1-p_1)^{1-y^{(i)}} + (1-\Phi_i)p_0^{y^{(i)}}(1-p_0)^{1-y^{(i)}}\}$. Let $R = -ψ^{-1}E\ln\{Ap_1^y(1-p_1)^{1-y} + (1-A)p_0^y(1-p_0)^{1-y}\}$.

(C2R): Let $R' = E\rho(y, A)$ where $A = I[x^Tβ > 0]$, $y \in \{0,1\}$, $\rho(0,0) < \rho(0,1)$ and $\rho(1,1) < \rho(1,0)$. Define $q = \rho(1,0) + \rho(0,1) - \rho(1,1) - \rho(0,0)$, and $h = [\rho(0,1) - \rho(0,0)]/q$.

(Rs): $R_n$ is an i.i.d. average of terms that are bounded between $[0, Q]$ for some positive constant $Q$.

(Rp): $R$ is nonstochastic and bounded between $[0, Q]$.

(C2L): (Uniform continuity for $R$.) There exists a constant $W > 0$ and a constant $\varepsilon > 0$ such that

$$|R(\beta) - R(\beta')| \leq W|\beta - \beta'|_1,$$

for all $\beta$ and $\beta'$ in $\Omega_n \subset \Re^{K_n}$, whenever $|\beta - \beta'|_1 \leq \varepsilon$.

(L): (Lipshitz for $R_n$.) For some $q' \geq 0$, for all large enough $n$,

$$|R_n(\beta) - R_n(\beta')| \leq n^{q'}|\beta - \beta'|_\infty$$

with probability 1, for all $\beta$ and $\beta'$ in $\Re^{K_n}$.

(B): (Bias.) $\sup_{\beta \in \Omega_n} |E_{D_n} R_n(\beta) - R(\beta)| \prec \delta_n$.

(C): $H_n$ is such that $\pi[R - \inf R(H_n) < \delta_n] \geq e^{-n\psi\delta_n}$ for all large enough $n$.

(C2b): $H_n$ $(\subset \Omega_n)$ is a compact set of $\beta$'s each satisfying the following: $\beta \in \Omega_n$, and for any small enough $\eta > 0$, there exists a large enough $N_\eta$, such that the prior $\pi$ around a neighborhood of $\beta$ satisfies $\pi[b : |b - \beta|_1 < \eta\delta_n] \geq e^{-n\psi\delta_n}$ for all $n > N_\eta$.

(T1): For some $M > 1$ and $u \geq 0$,

$$\pi(\Theta_n^c) \prec e^{-2n\psi c'}$$

for any constant $c' > 0$, where $\Theta_n^c = \Omega_n - \Theta_n$, $\Omega_n$ is the support of $\pi$, and the set $\Theta_n = \{\beta \in \Omega_n : |\beta|_0 (= |\gamma|_1) \leq \lceil Mn\delta_n^2/(\log n)^2 \rceil, |\beta|_\infty \leq n^u\}$.

$0'$. $|x_j| \leq 1$ for all $j$.

$0''$. The conditional density $p(x_1|\tilde{x})$ with respect to the Lebesgue measure exists for all $x$ and is bounded above by a constant $S > 0$.

$1'$. $1 \succ \delta_n \succ n^{-1/2}\log n$.

$3'$. $n \prec K \prec n^\alpha$ for some $\alpha > 1$.

LEMMA 1. $R'(\beta) = R(\beta) + c$ where $R$ is the risk function in (RnR), $R'$ is the risk function in (C2R), and $c$ is a constant. Both $R'$ and $R$ are equal to $qE(A(h - Y)) + a$ constant.

PROOF. Note that for all $a, y \in \{0, 1\}$, $\ln\{Ap_1^y(1 - p_1)^{1-y} + (1 - A)p_0^y(1 - p_0)^{1-y}\} = Ay\ln\{p_1(1 - p_0)p_0^{-1}(1 - p_1)^{-1}\} + A\ln\{(1 - p_1)(1 - p_0)^{-1}\} + y\ln\{p_0(1 - p_0)^{-1}\} + \ln(1 - p_0)$, which is, using the definitions of $p_{0,1}$, $Ay\psi q - Ah\psi q$ plus something that does not depend on $A$. Then due to (RnR), $R(\beta) = -\psi^{-1}E(Ay\psi q - Ah\psi q) = qE[A(h - y)]$, up to an additive constant.

In (C2R), $R'(\beta) = E\rho(Y, A) = \sum_{y,a \in \{0,1\}} \rho(y, a)E[Y^y(1 - Y)^{1-y}A^a(1 - A)^{1-a}] = \rho(0, 0) + (\rho(1, 1) + \rho(0, 0) - \rho(1, 0) - \rho(0, 1))EYA + (\rho(1, 0) - \rho(0, 0))EY + (\rho(0, 1) - \rho(0, 0))EA = \rho(0, 0) + (\rho(1, 0) - \rho(0, 0))EY + (\rho(1, 1) + \rho(0, 0) - \rho(1, 0) - \rho(0, 1))E[A(Y + (\rho(0, 1) - \rho(0, 0))/(\rho(1, 1) + \rho(0, 0) -$

$\rho(1,0) - \rho(0,1))] = constant + (\rho(1,0) + \rho(0,1) - \rho(1,1) - \rho(0,0)) \times E[A((\rho(0,1) - \rho(0,0))/(\rho(1,0) + \rho(0,1) - \rho(1,1) - \rho(0,0)) - Y)] = constant + qE[A(h-Y)]$, where $q = \rho(1,0) + \rho(0,1) - \rho(1,1) - \rho(0,0) > 0$ and $h = (\rho(0,1) - \rho(0,0))/(\rho(1,0) + \rho(0,1) - \rho(1,1) - \rho(0,0)) \in (0,1)$ due to $\rho(0,1) > \rho(0,0)$ and $\rho(1,0) > \rho(1,1)$. $\square$

REMARK 1. The risk function $R'$ in condition (C2R) [or equivalently $R$ in (RnR)] describes a risk function in data mining that is more general than the classification error. For one example in a data mining context: A marketing effort $A = I[\text{mail}]$ of mailing out an advertisement with cost $c = 1$ can be based on $x$ (including, e.g., gender, age, ethnic group, education, . . .) through a decision rule $A = I(x^T\beta > 0)$. The outcome will be $Y = I[\text{purchase}]$ where a purchase will lead to net income $g = 100$. Then one would like to maximize the expected profit $E[(gY-c)A]$ or minimize a risk $R = constant - E[(gY-c)A]$. Here up to a constant, $\rho(Y,A) = -(gY-c)A$, so that $\rho(0,0) = \rho(1,0) = 0$, $\rho(0,1) = c = 1$, $\rho(1,1) = c-g = -99$. Such profit-and-loss decision matrices are included in popular data mining software such as SAS Enterprise Miner. When $\rho(0,1) = \rho(1,0) = 1$ and $\rho(1,1) = \rho(0,0) = 0$, we obtain the special case of $R$ being the classification error, in which case $q = 2$ and $h = 0.5$.

REMARK 2. Consider the smooth sample risk used for classification [choice (ii) in Section 2]: $R_n = -\psi^{-1}n^{-1}\sum_{i=1}^{n}\log\{\Phi_i e^{\psi(y^{(i)}-1)} + (1-\Phi_i)e^{-\psi y^{(i)}}\}$, where $\Phi_i = \Phi(\sigma_n^{-1}(x^{(i)})^T\beta)$, $\Phi$ is the standard normal cumulative density function and $\sigma_n$ is a scaling factor.

It is noted that up to a constant of $\beta$, $e^{-n\psi R_n} = (constant) \times \prod_{i=1}^{n}\{\Phi_i p_1^{y^{(i)}}(1-p_1)^{1-y^{(i)}} + (1-\Phi_i)p_0^{y^{(i)}}(1-p_0)^{1-y^{(i)}}\}$, where $p_1 = e^{\psi}/(1+e^{\psi})$ and $p_0 = 1/(1+e^{\psi})$. So this forms a special case of $R_n$ in (RnR) with $h = 1/2$ and $q = 2$.

PROPOSITION 4 (General prior). *Under (Rs), (Rp), (C2L), (L), (B), (C2b), (T1), 1′, 3′, we have $R - \inf R(H_n) \le 6\delta_n$ with $P_{\beta,D}$-probability tending to 1 as $n \to \infty$, and $E_{\beta,D}R - \inf R(H_n) \le 6\delta_n$ for all large enough $n$. The same results hold if risk $R$ is replaced by a translated new risk $R' = R + c$ for any constant $c$.*

PROOF. The proposition is proved by combining Lemmas 2 and 3 below. $\square$

LEMMA 2. *Under (Rs), (Rp), (L), (B), (C) (in Proposition 7 of Section 6.1), (T1), 1′, 3′, we have $R - \inf R(H_n) \le 6\delta_n$ with $P_{\beta,D}$-probability tending* *to 1 as $n \to \infty$, and $E_{\beta,D} R - \inf R(H_n) \leq 6\delta_n$ for all large enough $n$. The same results hold if risk $R$ is replaced by a translated new risk $R' = R + c$ for any constant $c$.*

PROOF. We will apply Proposition 7 from Section 6.1. Here we can take $\bar{R} = Q$ and $\bar{R} - \inf R(H_n) \in [0, Q]$, due to (Rp). The parameter $b$ is denoted as $\beta$ here. The set $F_n$ is denoted as $\Theta_n$ here. Note that $R_n \geq 0$ due to (Rs), $\delta_n \prec 1$ due to condition 1'. Note that (T1) implies (T).

To prove the bounds in Lemma 2, it suffices to show that the two terms on the right-hand side of (1) in Proposition 7 are both $o(\delta_n)$. The second term $4e^{-n\psi\delta_n} \prec \delta_n$ due to the condition 1' for $\delta_n$. The first term $P^*[\sup_{\beta \in \Theta_n} |R_n(\beta) - R(\beta)| > \delta_n]$ is bounded above by $P^*[\sup_{\beta \in \Theta_n} |R_n(\beta) - ER_n(\beta)| > \delta_n/2]$ for all large $n$, due to the bias condition (B) $\sup_{\beta \in \Theta_n} |E_{D_n} R_n(\beta) - R(\beta)| \prec \delta_n$. Therefore it suffices to prove that $P^*[\sup_{\beta \in \Theta_n} |R_n(\beta) - ER_n(\beta)| > \delta_n/2] \prec \delta_n$.

Note that in general, due to (Rs), $P^*[\sup_{b \in \Theta_n} |R_n(b) - ER_n(b)| > \varepsilon_n]$ can be bounded using a covering number for $\Theta_n$, a Lipschitz condition $|R_n(b) - R_n(b')| \leq L_n |b - b'|_\infty$, a union bound and a Hoeffding inequality.

Suppose $\Theta_n$ can be covered by $N$ balls of radius $s$, such that for any $b \in \Theta_n$, there exists a $b_k \in \Re^{K_n}$, $k \in \{1, \ldots, N\}$, such that $|b - b_k|_\infty < s$. Then for any $b \in \Theta_n$, one can find one of these $N$ $b_k$'s, say, $b_j$, such that $|R_n(b) - ER_n(b)| - |R_n(b_j) - ER_n(b_j)| \leq 2\varepsilon_n/3$ by choosing $s = \varepsilon_n/(3L_n)$, due to the Lipschitz condition. Therefore $\sup_{b \in \Theta_n} |R_n(b) - ER_n(b)| \leq \sup_{j \in \{1, \ldots, N\}} |R_n(b_j) - ER_n(b_j)| + 2\varepsilon_n/3$. Then $P^*[\sup_{b \in \Theta_n} |R_n(b) - ER_n(b)| > \varepsilon_n] \leq P^*[\sup_{j \in \{1, \ldots, N\}} |R_n(b_j) - ER_n(b_j)| > \varepsilon_n/3]$ which is at most $N2e^{-2n(\varepsilon_n/3)^2 Q^{-2}}$ due to the union bound, condition (Rs), and the Hoeffding inequality.

Note that one can choose $N \leq \bar{N} = \sum_{r \leq d} K^r (n^u/s + 1)^r$, where $d = \lceil Mn\delta_n^2 / (\log n)^2 \rceil$, since the definition of $\Theta_n$ implies that there can be at most $K^r$ "model" indicator $\gamma$'s with size $|\gamma|_1 = r$ ($r \leq d$), each of which has a parameter space (of the nonzero $\beta$-components) that can be covered by at most $(2n^u/(2s) + 1)^r$ balls of size $s$. Now $\sum_{r \leq d} K^r (n^u/s + 1)^r \leq (d+1) K^d (n^u/s + 1)^d \leq K^{d+1}(n^u/s + 1)^d \leq K^{2d}(n^u/s + 1)^d \leq n^{2d\alpha}(n^u/s + 1)^d$ for all large $n$, since $1 \prec d \prec K \prec n^\alpha$ due to $1 \prec d \prec n$ (implied by condition 1') and $n \prec K \prec n^\alpha$ (by condition 3'). So we can choose $N \leq n^{2d\alpha}(n^u/s + 1)^d$ for all large enough $n$, where $s = \varepsilon_n/(3L_n)$ as prescribed before.

Now taking $\varepsilon_n = \delta_n/2$, $L_n = n^{q'}$ [from condition (L)], we get, for all large $n$,

$$P^*\left[\sup_{b \in \Theta_n} |R_n(b) - ER_n(b)| > \delta_n/2\right]$$
$$\leq \left[n^{2\alpha}\left(\frac{n^u}{(\delta_n/2)/(3n^{q'})} + 1\right)\right]^{\lceil Mn\delta_n^2/(\log n)^2 \rceil} 2e^{-2n(\delta_n/2/3)^2 Q^{-2}},$$

which can be proved to be less than $\delta_n$ in order due to the condition 1′: $1 \succ \delta_n \succ n^{-1/2} \log n$.

Collecting these steps together leads to the proof. $\square$

LEMMA 3. *If the conditions (C2L) and (C2b) are satisfied for some sequence $\delta_n \prec 1$, then (C) is also satisfied for the same $\delta_n$.*

PROOF. Note that $\inf R(H_n)$ is achieved at some $\beta \in H_n$ (possibly depending on $n$) due to the compactness of $H_n$ in (C2b) and the continuity of $R$ implied by (C2L). Then for any $b \in \Omega_n$, we have (*) $R(b) - \inf R(H_n) = R(b) - R(\beta) \leq |R(b) - R(\beta)| \leq W|b - \beta|_1$ for all small enough $|b - \beta|_1$. Therefore for any sequence $\delta_n \prec 1$, $|b - \beta|_1 < \delta_n W^{-1}$ implies $R(b) - \inf R(H_n) < W\delta_n W^{-1} = \delta_n$, for all large $n$. Therefore $\pi[b : R(b) - \inf R(H_n) < \delta_n] \geq \pi[b : |b - \beta|_1 < \delta_n W^{-1}]$, which is $\geq e^{-n\psi\delta_n}$ for all large $n$, by taking $\eta = W^{-1}$ in (C2b). $\square$

REMARK 3. Note that (C2b) is a condition for proving (C) (see Lemma 3 above). Condition (C) describes that the prior $\pi$ is competitive against the rules in $H_n$ in some sense [when comparing the generated $R(\beta)$'s to $\inf R(H_n)$]. Condition (C2b) describes one way to construct such a set of rules $H_n$ over which the prior $\pi$ is competitive: a compact set of rules such that around each of these rules the prior assigns a not too low probability.

LEMMA 4. (i) *Condition (RnR) implies (Rs) and (Rp).*

(ii) *Conditions 0′, 0″ and (RnR) imply (C2L).*

(iii) *Conditions 0′, $(\sigma)$, 3′ and (RnR) imply (L).*

(iv) *Conditions 0″, 1′, $(\sigma)$ and (RnR) imply (B).*

PROOF. For (i), note that the proofs for (Rs) and (Rp) are similar. Note that $h \in (0,1)$ and $q > 0$ implies that $p_{0,1} \in (0,1)$. The terms inside $\ln\{\ \}$ are averages of $p_1$ and $p_0$ or averages of $(1-p_1)$ and $(1-p_0)$, which are all between $(min, 1)$, where $min = \min\{p_{0,1}, (1-p_{0,1})\} \in (0,1)$. This implies that $-\psi^{-1} \ln\{\ \} \in (0, \psi^{-1} \ln(1/min))$. So (Rs) and (Rp) are proved with $Q = \psi^{-1} \ln(1/min)$.

For (ii), note that $R = qE[A(h-y)]$, up to an additive constant, due to Lemma 1.

For any $b$ and $\beta$ in $\{\pm 1\} \times \Re^{K-1}$ such that $|b - \beta|_1 \leq \varepsilon$ for some small enough $\varepsilon$, we must have $b_1 = \beta_1 \in \{\pm 1\}$. Let us take $b_1 = \beta_1 = +1$. (The other case is similar.) Then $|R(b) - R(\beta)| = |qE[(I[x^T b > 0] - I[x^T \beta > 0]])(h - y)]| \leq qE|I[x^T \beta > 0] - I[x^T b > 0]|$. Here we are using the representations such as $b^T = (b_1, \tilde{b}^T)$ and $\beta^T = (\beta_1, \tilde{\beta}^T)$.

Now $b_1 = \beta_1 = 1$ implies that ($\ddagger$) $E|I[x^T\beta > 0] - I[x^Tb > 0]| = E_{\tilde{x}}E_{x_1|\tilde{x}} \times I[-\tilde{x}^T\tilde{b} \geq x_1 > -\tilde{x}^T\tilde{\beta} \text{ or } -\tilde{x}^T\tilde{\beta} \geq x_1 > -\tilde{x}^T\tilde{b}] \leq ES|\tilde{x}^T\tilde{b} - \tilde{x}^T\tilde{\beta}| \leq ES \times |x|_\infty|b - \beta|_1 \leq S|b - \beta|_1$, where $|x|_\infty \leq 1$ due to $0'$ and $S$ is an upperbound of the conditional density of $p(x_1|\tilde{x})$ in $0''$. So $|R(b) - R(\beta)| \leq qE|I[x^T\beta > 0] - I[x^Tb > 0]| \leq qS|b - \beta|_1$. So we can take $W = qS$ to obtain the proof for (ii).

For (iii), note that (*) $|R_n(b) - R_n(b')| \leq K_n C_n|b - b'|_\infty$, where $C_n$ is any upperbound of $|\partial_{b_j} R_n|$ over all $j$ and over parameter space. Note that $|\partial_{b_j} R_n| = |-(n\psi)^{-1}\sum_{i=1}^n \{\Phi_i p_1^{y^{(i)}}(1-p_1)^{1-y^{(i)}} + (1-\Phi_i)p_0^{y^{(i)}}(1-p_0)^{1-y^{(i)}}\}^{-1}(\partial_{b_j}\Phi_i) \times \{p_1^{y^{(i)}}(1-p_1)^{1-y^{(i)}} - p_0^{y^{(i)}}(1-p_0)^{1-y^{(i)}}\}| \leq (n\psi)^{-1}\sum_{i=1}^n (min)^{-1}(1/\sqrt{2\pi})\sigma_n^{-1} \times 1$, since $|p_1^{y^{(i)}}(1-p_1)^{1-y^{(i)}} - p_0^{y^{(i)}}(1-p_0)^{1-y^{(i)}}| \leq 1$, $\{\Phi_i p_1^{y^{(i)}}(1-p_1)^{1-y^{(i)}} + (1-\Phi_i)p_0^{y^{(i)}}(1-p_0)^{1-y^{(i)}}\} \geq min = \min\{p_{0,1}, (1-p_{0,1})\} \in (0,1)$, and $|\partial_{b_j}\Phi_i| = |\partial_{b_j}\Phi(x^{(i)T}b/\sigma_n)| = |x_j^{(i)}|\sigma_n^{-1}(1/\sqrt{2\pi})e^{-0.5(x^{(i)T}b/\sigma_n)^2} \leq \sigma_n^{-1}(1/\sqrt{2\pi})$ due to $0'$.

So $|\partial_{b_j}R_n| \leq \psi^{-1}(min)^{-1}(1/\sqrt{2\pi})\sigma_n^{-1}$, which can be taken as the upperbound $C_n$ in (*), which implies that $|R_n(b) - R_n(b')| \leq (constant)K_n\sigma_n^{-1}|b - b'|_\infty \leq n^{\alpha+q''+1}|b - b'|_\infty$ for all large $n$, due to conditions $3'$ and ($\sigma$). Then (C2L) is proved with $q' = \alpha + q'' + 1$.

For (iv), note that $|ER_n - R| = \psi^{-1}|E\ln\{\Phi p_1^y(1-p_1)^{1-y} + (1-\Phi)p_0^y(1-p_0)^{1-y}\} - E\ln\{Ap_1^y(1-p_1)^{1-y} + (1-A)p_0^y(1-p_0)^{1-y}\}|$, where $\Phi = \Phi(x^T\beta/\sigma_n)$ and $A = I(x^T\beta > 0)$. By a first-order Taylor expansion one shows that $|ER_n - R| \leq constant E|\Phi - A| = constant E_{\tilde{x}}E_{x_1|\tilde{x}}|\Phi - A|$. Now suppose $\beta_1 = 1$ (the case of $\beta_1 = -1$ is similar); then $E_{x_1|\tilde{x}}|\Phi - A| = E_{x_1|\tilde{x}}\{I(|x_1 + \tilde{x}^T\tilde{\beta}| \leq u)|\Phi - A|\} + E_{x_1|\tilde{x}}\{I(|x_1 + \tilde{x}^T\tilde{\beta}| > u)|\Phi - A|\} \leq S(2u) + e^{-0.5(u/\sigma_n)^2}/\sqrt{2\pi(u/\sigma_n)^2}$ for any $u > 0$, due to $0''$ and the Mill's ratio. The resulting upperbound is uniformly correct for all $\beta$, and becomes $O(\log n/\sqrt{n})$ by taking $u = \sigma_n\sqrt{\log n}$ and using ($\sigma$).

So $\sup_\beta |ER_n - R| \leq O(\log n/\sqrt{n}) \prec \delta_n$ due to $1'$.  $\square$

LEMMA 5.  (i) $H_1 \supset H_2$.

  (ii) $H_2 \supset H_3$ for all large $n$.
  (iii) $H_2 \supset H_b$ for all large $n$.
  (iv) $H_3 \not\supset H_b$ assuming condition $1'$.
  (v) $H_b \not\supset H_3$ assuming $1'$ and $3'$.
  (vi) $H_m \supset H_E$ for all large enough $q$.
  (vii) $H_3 \supset H_m$ for all large $n$ if $\delta_n \geq n^{-m/(2m+1)}(\log n)^2$, assuming $1'$.
  (viii) $H_3 \supset H_E$ for all large $n$ if $\delta_n \geq n^{-1/2}(\log n)^2$, assuming $1'$.
  (ix) $H_b \not\supset H_E$ under $3'$, for all large enough $q$.

PROOF.  Part (vi) is due to the domination of power law decays over the exponential decays. Parts (vii) and (viii) can be proved by applying the (polynomial and exponential, resp.) bounds of $\sum_{j>r} |\tilde{\beta}_{(j)}|$ for $r = v_n \equiv n\delta_n^2/(\log n)^2$. Part (iv) is proved by examining $\beta = (1, C, \ldots, C, 0, \ldots, 0)^T$ with about $v_n$ $C$'s, which is a number of $H_b$ but not of $H_3$ due to an $\ell_1$ norm that is unbounded as $n$ increases. Part (v) is proved by examining $\beta = (1, (1/2)C'\delta_n/(\log n), (1/2)^2C'\delta_n/(\log n), (1/2)^3C'\delta_n/(\log n), \ldots)^T$ which is a member of $H_3$ but not of $H_b$. Part (iii) is proved by noting that $H_b$ implies a zero tail for the sum over $j > v_n$ and bounded terms for $j \leq v_n$. Part (ii) is proved by noting that a bounded $\ell_1$ norm implies that all coefficients are bounded. Part (i) is proved by noting that $\sum_{j \leq v_n} |\tilde{\beta}_{(j)}|^2 \leq (\sup_{j \leq v_n} |\tilde{\beta}_{(j)}|)^2 v_n$. To prove part (ix), note that $\beta = (1, \psi_0\xi, \psi_0\xi^2, \psi_0\xi^3, \ldots)^T$ is a member of $H_E$ for all large $q$, if $\psi_0 = C(e^{2C''} - 1)$, $\xi = e^{-2C''}$. On the other hand $\beta \notin H_b$ under $3'$. □

LEMMA 6. (i) $R - \inf(H_{2,3,b}) \leq R - \inf R(H_1)$.

(ii) *If* $\delta_n \geq n^{-m/(2m+1)}(\log n)^2$, *then* $R - \inf(H_m) \leq R - \inf R(H_1)$.

(iii) *If* $\delta_n \geq n^{-1/2}(\log n)^2$, *then* $R - \inf(H_E) \leq R - \inf R(H_1)$.

PROOF. Note that the previous lemma on the relations among the sparse sets implies that $H_1$ contains all other sparse sets in all situations of this lemma [with the specifications of $\delta_n$ for situations (i) and (iii)]. Then $\inf R(H_1)$ is the smallest among all these infimums and $R - \inf R(H_{2,3,b,m,E}) \leq R - \inf R(H_1)$. □

PROPOSITION 5. *Propositions 2 and 3 hold for the more general risk functions in a data mining context specified in (RnR).*

PROOF. We only need prove the Scenario-II results, since they obviously imply the Scenario-I results. [If we assume that $\inf_{\beta \in \Omega_n} R(\beta)$ is achieved at some $\beta_H \in H_n \subset \Omega_n$, then $\inf_{\beta \in \Omega_n} R(\beta) = \inf_{\beta \in H_n} R(\beta)$.]

Due to Lemma 6(i) it is obvious that we only need to prove Proposition 3 for $H_1$, in order to prove Proposition 3. For getting the upperbounds of the risk performances in Proposition 2 we start from Proposition 3 and apply Lemma 6(ii) and (iii), where we take "$=$" for the choices of $\delta_n$, and note that the factors $(\log n)^2$ in $\delta_n$ are less than the factor $n^\xi$ for all large enough $n$, for any $\xi > 0$.

To prove Proposition 3 for $H_1$ in the general context (RnR), we apply Proposition 4 and note that all conditions hold, by applying Lemma 4, as well as Lemmas 7 and 8 to be given later. □

LEMMA 7. *For $\delta_n$ in condition $1'$, assume that $K_n$ satisfies $3'$ and that we use the normal/binary prior for $\pi$ satisfying (V). Assume condition $(r_\delta)$. Then the sparse set $H_1$ satisfies the condition (C2b).*

PROOF.   The set $H_1$ is obviously compact.

Consider any $\beta \in H_1$. Define "model" $\gamma_n$ to be the set of indices for the $(\beta_j)_{j>1}$'s that have the top $\lceil v_n \rceil$ largest absolute values, in addition to the index 1 for $\beta_1 \in \{\pm 1\}$ which is always kept in the "model." Then $\sum_{j \notin \gamma_n} |\beta_j| = \sum_{j > v_n} |\check{\beta}_{(j)}| \le C' \delta_n / (\log n)$. Here $v_n = n \delta_n^2 / (\log n)^2$.

Note that for any $\eta > 0$, $\pi[b \colon |b - \beta|_1 < \eta \delta_n] \ge \pi(\gamma = \gamma_n) \pi[b \colon |b - \beta|_1 < \delta_n \eta | \gamma = \gamma_n]$, where $\gamma$ is the "model" indicator for the set of nonzero components of $b$. Following the notation of Section 4.2, $\gamma = (\gamma_j)_1^{K_n}$, where $\gamma_j = I(|b_j| \ne 0)$. Here and below, for a $K_n$-vector $\zeta$ (e.g., $\zeta$ can be $\gamma$ or $\check{\gamma}$), the notation $\zeta = \gamma_n$ for a set $\gamma_n \subset \{1, \ldots, K_n\}$ means that $\zeta_j = I[j \in \gamma_n]$, $j = 1, \ldots, K_n$.

We will show that $\pi(\gamma = \gamma_n)$ and $\pi[b \colon |b - \beta_1| < \delta_n \eta | \gamma = \gamma_n]$ are both not too small.

Note that given "model" $\gamma_n$, $b_1 = \beta_1$ and $|\tilde{b}_\gamma - \tilde{\beta}_\gamma|_1 \equiv \sum_{j>1, j \in \gamma_n} |b_j - \beta_j| < \delta_n / \log n$ will imply that $|b - \beta|_1 < \delta_n \eta$ (for all large enough $n$). This is because $b$ only has nonzero components in $\gamma_n$, so $|b - \beta|_1 = |b_1 - \beta_1| + \sum_{j>1, j \in \gamma_n} |b_j - \beta_j| + \sum_{j \notin \gamma_n} |\beta_j| \le 0 + \delta_n / \log n + C' \delta_n / (\log n)$ which is less than $\delta_n$ in order.

So $\pi[b \colon |b - \beta|_1 < \delta_n \eta | \gamma = \gamma_n] \ge \pi[b_1 = \beta_1, |\tilde{b}_\gamma - \tilde{\beta}_\gamma|_1 < \delta_n / \log n | \gamma = \gamma_n] = 0.5 \pi[|\tilde{b}_\gamma - \tilde{\beta}_\gamma|_1 < \delta_n / \log n | \gamma = \gamma_n]$ for all large $n$, noting that $b_1 \in \pm 1$ with equal probability and is independent of other things in the prior. The last probability $\pi[|\tilde{b}_\gamma - \tilde{\beta}_\gamma|_1 < \delta_n / \log n | \gamma = \gamma_n]$ is integrating a normal density $|2\pi V_\gamma|^{-1/2} e^{-0.5 \tilde{b}_\gamma^T V_\gamma^{-1} \tilde{b}_\gamma}$ over a set $S = [|\tilde{b}_\gamma - \tilde{\beta}_\gamma|_1 < \delta_n / \log n] \supset [v_n | \tilde{b}_\gamma - \tilde{\beta}_\gamma|_\infty < \delta_n / \log n]$, which has at least volume $(v_n^{-1} \delta_n / \log n)^{\lceil v_n \rceil}$ since the vector $\tilde{b}_\gamma$ is $\lceil v_n \rceil$-dimensional under model $\gamma = \gamma_n$.

The normal density over is bounded below by $\exp\{-0.5 v_n \log(2\pi B) - 0.5 |\tilde{b}_\gamma|_2^2 B\}$ using the bounds of the eigenvalues of prior variance in $(V)$. Note also that $|\tilde{b}_\gamma|_2^2 \le 2|\tilde{\beta}_\gamma|_2^2 + 2|\tilde{b}_\gamma - \tilde{\beta}_\gamma|_2^2 \le 2|\tilde{\beta}_\gamma|_2^2 + 2|\tilde{b}_\gamma - \tilde{\beta}_\gamma|_1^2 \le 2|\tilde{\beta}_\gamma|_2^2 + 2\delta_n^2 / (\log n)^2$ over $\tilde{b}_\gamma \in S$, which is $\le 2C^2 n \delta_n^2 / (\log n) + 2\delta_n^2 / (\log n)^2$ since $\beta \in H_1$.

Collecting all these together we get, for all large $n$,

$$\pi[|b - \beta|_1 < \delta_n \eta | \gamma = \gamma_n]$$
$$\ge 0.5 \exp\{-0.5 v_n \log(2\pi B) - C^2 n \delta_n^2 / (\log n) B$$
$$- \delta_n^2 / (\log n)^2 B\}(v_n^{-1} \delta_n / \log n)^{\lceil v_n \rceil}\}$$
$$= 0.5 \exp\{-0.5 v_n \log(2\pi B) - C^2 n \delta_n^2 / (\log n) B$$
$$- \delta_n^2 / (\log n)^2 B - \lceil v_n \rceil \log(v_n \log n / \delta_n)\},$$

where $v_n = n \delta_n^2 / (\log n)^2$. It is then easy to verify that all terms in the exponent are of the form $-o(n \delta_n^2)$ under condition 1' for $\delta_n$.

Now we consider $\pi(\gamma = \gamma_n)$ under the (size-restricted) binary prior. Note that for all large enough $n$, $\pi(\gamma = \gamma_n) = \pi(\check{\gamma} = \gamma_n || \check{\gamma}|_1 \leq \bar{r}_n) \geq \pi(\check{\gamma} = \gamma_n, |\check{\gamma}|_1 \leq \bar{r}_n) = \pi(\check{\gamma} = \gamma_n)$, where $\bar{r}_n$ is the size restriction chosen as (the integer part of) $Mn\delta_n^2/(\log n)^2$ ($M > 1$) in condition $(r_\delta)$, and the "model" $\gamma_n$ has size $1 + \lceil v_n \rceil = \lceil n\delta_n^2/(\log n)^2 \rceil + 1 < \bar{r}_n$ for all large enough $n$. Note that $\check{\gamma}$ has unrestricted i.i.d. binary components (except that $\check{\gamma}_1 = 1$ always) and the probability $\pi(\check{\gamma} = \gamma_n) = \lambda_n^{\lceil v_n \rceil}(1 - \lambda_n)^{K_n - 1 - \lceil v_n \rceil}$. Note that $\lambda_n \sim v_n/K_n$ due to $(r_\delta)$ and $v_n \prec K_n$ due to $1'$ and $3'$. Therefore $\log \pi(\check{\gamma} = \gamma_n) = \lceil v_n \rceil \log \lambda_n + (K_n - 1 - \lceil v_n \rceil) \log(1 - \lambda_n) = \lceil v_n \rceil \log \lambda_n + (K_n - 1 - \lceil v_n \rceil)(-\lambda_n + o(\lambda_n)) = \lceil v_n \rceil \log \lambda_n(1 + o(1)) \geq (\lceil v_n \rceil \log(M'v_n) - \lceil v_n \rceil \log K_n)(1 + o(1)) \geq -v_n \log K_n(1 + o(1))$ for all large $n$ [since $v_n = n\delta_n^2/(\log n)^2 \succ 1$ due to $1'$]. Now $v_n \log K_n = n\delta_n^2/(\log n)^2 \log K_n \leq [n\delta_n^2/(\log n)^2] \log(n^\alpha) = o(n\delta_n^2)$.

Collecting these results together we know that $\pi[b : |b - \beta|_1 < \eta\delta_n] \geq \pi(\gamma = \gamma_n)\pi[b : |b - \beta|_1 < \delta_n\eta|\gamma = \gamma_n]$ where both factors can be expressed as being at least $e^{-o(n\delta_n^2)}$, which will be greater than $e^{-\psi n\delta_n}$, for all large $n$. (Note that $\delta_n \prec 1$ due to $1'$ is used.) □

LEMMA 8. *With conditions (V), $(r_\delta)$, $1'$, the normal-binary prior $\pi$ (with size restriction) satisfies the tail condition (T1).*

PROOF. Take $M$ as the one used in condition $(r_\delta)$ and take $u = 1$ in (T1). Denote $\bar{r}_n = \lceil Mn\delta_n^2/(\log n)^2 \rceil$. Note that $\pi(\Theta_n^c) \leq \pi(|\gamma|_1 > \bar{r}_n) + \sum_{\gamma : |\gamma|_1 \leq \bar{r}_n} \pi[|\beta|_\infty > n^u|\gamma]\pi(\gamma) \leq \pi(|\gamma|_1 > \bar{r}_n) + \sup_{\gamma : |\gamma|_1 \leq \bar{r}_n} \pi[|\beta|_\infty > n^u|\gamma]$. The first term is 0 due to the size restriction. The term $\sup_{\gamma : |\gamma|_1 \leq \bar{r}_n} \pi[|\beta|_\infty > n^u|\gamma] = \sup_{\gamma : |\gamma|_1 \leq \bar{r}_n} \pi[\bigcup_{j : \gamma_j = 1} [|\beta_j| > n^u]|\gamma] \leq \bar{r}_n \sup_{\gamma : |\gamma|_1 \leq \bar{r}_n} \sup_{j : \gamma_j = 1} \pi[|\beta_j| > n^u|\gamma]$, where $\pi[|\beta_j| > n^u|\gamma]$ can be bounded above by $2e^{-0.5n^{2u}/B}/\sqrt{2\pi n^{2u}/B}$ using Mill's ratio and the eigenvalue bound (V). Collecting all these together we get $\pi(\Theta_n^c) \leq \bar{r}_n 2e^{-0.5n^{2u}/B}/\sqrt{2\pi n^{2u}/B}$ (where $u$ is taken to be 1), which is at most $e^{-0.5n^2/B}$ under conditions $(r_\delta)$ and $1'$, for all large $n$, and is therefore $\prec e^{-2n\psi c'}$ for any constant $c' > 0$. □

6.1. *Supplementary results on risk performance of Gibbs posterior.* In this section we will consider a very general setup. The results here have been applied in the proofs in Section 6. Here we will consider the performance of a general risk $R(b)$ [or more generally, $r_n(b)$, which is nonstochastic but can depend on $n$]. Suppose $b$ is sampled from a Gibbs posterior $\omega(db|D_n)$, which is constructed from a sample risk $R_n$ and a prior $\pi(db)$, and $D_n$ denotes data generated from a true density $p^*$.

More formally, in both propositions below, we will assume that the data $D_n$ (indexed by a sample size $n$) follows a probability distribution $P^*$ with density $p^*(D_n)$ with respect to some dominant measure $dD_n$. Let $b|D_n$ denote a distribution (conditional on $D_n$) with a density $w(b|D_n) \propto e^{-n\psi R_n(b)}$ with respect to a prior $\pi(db)$, where $R_n(b)$ depends on a parameter $b$ and data $D_n$. Denote by $P_{b,D}$ the resulting joint distribution of $b$ and $D_n$ and $E_{b,D}$ the corresponding expectation.

PROPOSITION 6.   *Assume that $R_n(b) \geq 0$ for any $b$ and $D_n$.*

*If the prior $\pi$ is such that the support $\operatorname{supp}(\pi) = \Omega_n = F_n \cup F_n^c$ where $F_n^c = \Omega_n - F_n$, then for any $r_n(b)$ nonstochastic (possibly depending on $n$ but not otherwise on $D_n$) and any $\rho_n$ and $\delta_n$ nonstochastic and not depending on $p$,*

$$P_{b,D}[r_n(p) - \rho_n > 5\delta_n] \leq P^*\left[\sup_{b \in F_n} |R_n(b) - r_n(b)| > \delta_n\right]$$

$$+ \frac{\pi(F_n^c)e^{n\psi(\rho_n + 2\delta_n)} + e^{-n\psi(2\delta_n)}}{(\pi[r_n(b) - \rho_n < \delta_n] - \pi(F_n^c))_+}.$$

*Here we use the notation $A_+ = AI(A > 0)$.*

PROOF.   The left-hand side is $E_D\Psi = \int P^*(dD_n)[\frac{N1+N2}{Den}]$, where $E_D = \int P^*(dD_n)$, $\Psi = [\frac{N1+N2}{Den}]$, $N1 = \int_{F_n^c} e^{-n\psi(R_n - \rho_n)} I[r_n - \rho_n > 5\delta_n]\pi(db)$, $N2 = \int_{F_n} e^{-n\psi(R_n - r_n + r_n - \rho_n)} I[r_n - \rho_n > 5\delta_n]\pi(db)$, $Den = \int e^{-n\psi(R_n - r_n + r_n - \rho_n)}\pi(db)$. Note that $N1 \leq \pi(F_n^c)e^{n\psi\rho_n}$, $N2 \leq e^{n\psi\Delta_n - n\psi(5\delta_n)}$, where $\Delta_n = \sup_{F_n} |R_n(b) - r_n(b)|$,

$$Den \geq \int_{F_n} I[r_n - \rho_n < \delta_n] e^{-n\psi\Delta_n - n\psi(r_n - \rho_n)}\pi(db)$$

$$\geq e^{-n\psi\Delta_n - n\psi\delta_n}\pi([r_n - \rho_n < \delta_n] \cap F_n)$$

$$\geq e^{-n\psi\Delta_n - n\psi\delta_n}(\pi[r_n - \rho_n < \delta_n] - \pi(F_n^c))_+.$$

Therefore $\Psi = [\frac{N1+N2}{Den}] \leq G(\Delta_n)$, where

$$G(\Delta_n) = \left[\frac{e^{(\Delta_n + \delta_n + \rho_n)n\psi}\pi(F_n^c) + e^{(\Delta_n + \delta_n + \Delta_n - 5\delta_n)n\psi}}{(\pi[r_n - \rho_n < \delta_n] - \pi(F_n^c))_+}\right].$$

Note that $\Psi = P_{b|D_n}[r_n - \rho_n > 5\delta_n] \leq 1$ and $G(\Delta_n)$ is increasing in $\Delta_n$. Then the left-hand side is

$$E_D\Psi = E_D(\Psi I[\Delta_n > \delta_n]) + E_D(\Psi I[\Delta_n \leq \delta_n])$$

$$\leq P^*[\Delta_n > \delta_n] + E_D\{G(\Delta_n)I[\Delta_n \leq \delta_n]\}$$

$$\leq P^*[\Delta_n > \delta_n] + G(\delta_n)$$

$$\leq P^*[\Delta_n > \delta_n] + \left[\frac{e^{(2\delta_n + \rho_n)n\psi}\pi(F_n^c) + e^{(-2\delta_n)n\psi}}{(\pi[r_n - \rho_n < \delta_n] - \pi(F_n^c))_+}\right]. \qquad \square$$

PROPOSITION 7. *Assume that $R_n(b) \geq 0$ for any $b$ and $D_n$. Consider any positive sequence $\delta_n$ which is nonstochastic and not dependent on $b$. Assume that $\delta_n \prec 1$. For all large enough $n$, if the prior $\pi$ is such that the support $\mathrm{supp}(\pi) = \Omega_n = F_n \cup F_n^c$ where $F_n^c = \Omega_n - F_n$, such that*

(T) $\pi(F_n^c) \leq e^{-2n\psi\bar{R}}$ *for some constant $\bar{R} > 0$,*

(C) *a subset $H_n$ of the $\mathrm{supp}(\pi)$ is such that $\pi[R(b) - \inf_{b \in H_n} R(b) < \delta_n] \geq e^{-n\psi\delta_n}$, for some nonstochastic $R(b) \leq \bar{R}$,*

*then we have, for all large enough $n$,*

(1)
$$P_{b,D}\left[R(b) - \inf_{b \in H_n} R(b) > 5\delta_n\right]$$
$$\leq P^*\left[\sup_{b \in F_n} |R_n(b) - R(b)| > \delta_n\right] + 4e^{-n\psi\delta_n}$$

*and*

(2)
$$E_{b,D}\left[R(b) - \inf_{b \in H_n} R(b)\right]$$
$$\leq 5\delta_n + \left(\bar{R} - \inf_{b \in H_n} R(b)\right) P_{b,D}\left[R(b) - \inf_{b \in H_n} R(b) > 5\delta_n\right].$$

PROOF. Note that the second inequality relates expectation $E$ to a probability $P$, which is bounded in the first inequality. Such a relation is due to the general relation for a constant $g > 0$ and a random variable $G$ which is bounded above by a constant $c$: $EG = EGI[G > g] + EGI[G \leq g] \leq cP[G > g] + g$. We can then simply take $g = 5\delta_n$ and $G = R(b) - \inf_{b \in H_n} R(b)$, which is bounded above by a constant $c = \bar{R} - \inf_{b \in H_n} R(b)$.

Now we prove the first inequality on $P$. This is proved by applying Proposition 6. We take $r_n(b) = R(b)$, $\rho_n = \inf_{b \in H_n} R(b)$, and apply the conditions (T) and (C) to the long fraction on the right-hand side of the inequality in Proposition 6, which is shown to be bounded above by $4e^{-n\psi\delta_n}$, by noting that $\bar{R} > 0$ and $\delta_n \prec 1$. □

REMARK 4. This proposition simplifies the long fraction in Proposition 6 by applying the conditions (T) and (C) on the prior $\pi$ and "a scope of comparison" $H_n$. Then the performance of $R(b)$ is evaluated under $E_{b,D}$ or $P_{b,D}$ (as generated by the data generation mechanism $P^*$ for $D_n$ and the Gibbs posterior for $b|D_n$). The performance of $R(b)$ is compared to the best performance $\inf_{b \in H_n} R(b)$ over the scope $H_n$. It will be close to this best performance if $n^{-1} \prec \delta_n \prec 1$ and if there exists a uniform convergence result for a small $P^*[\sup_{b \in F_n} |R_n(b) - R(b)| > \delta_n]$. Such a relation is very general and allows many different situations by invoking different techniques. For

example, Vapnik–Chervonenkis theory, or uniform continuity of $R_n(b) - R(b)$ and covering numbers of $F_n$, may be used to handle the $P^*[\sup \ldots]$ with a union bound. The probability of large $|R_n - R|$ may also be bounded by Hoeffding's or Bernstein's inequalities for non-i.i.d. data or data that are dependent in some weak ways (such as $\alpha$- or $\phi$-mixing, ergodic Markov chain, etc.).

**7. An MCMC algorithm.** This section describes some computational aspect for sampling from the Gibbs posterior $\omega(d\beta|D^n) \propto e^{-n\psi R_n}\pi(d\beta)$, where $\pi$ is the normal-binary prior specified in Sections 4.2 and 4.3.

Consider the smoothed sample risk function $R_n$ in (RnR). It is noted that

$$e^{-n\psi R_n} = \prod_{i=1}^{n} \{\Phi_i p_1^{y^{(i)}}(1-p_1)^{1-y^{(i)}} + (1-\Phi_i)p_0^{y^{(i)}}(1-p_0)^{1-y^{(i)}}\},$$

where $\Phi_i = \Phi(\sigma_n^{-1}(x^{(i)})^T\beta)$, $\Phi$ is the standard normal cumulative density function and $\sigma_n$ is a scaling factor.

This can be recognized as the likelihood for a mixture of two binary models with mixing probability $\Phi_i$. This suggests a data augmentation method [see, e.g., Tanner (1996)] incorporating latent variables $Z = (Z^{(i)})_1^n$, where $Z^{(i)}$ are independent $N((x^{(i)})^T\beta, \sigma_n)$, so that $y^{(i)}|Z^{(i)}$ are independent $Bin(1, p_{I[Z^{(i)}>0]})$, which leads to computational advantage. The Gibbs sampler can be used to obtain the joint distribution of $(Z, \gamma, \beta)$, where all full conditional distributions are standard. Similar to, for example, Lee et al. (2003), we can integrate over $\beta_\gamma$ and use the distribution $\gamma|Z$ instead of $\gamma|Z, \beta_\gamma$ in the Gibbs sampler, in order to speed up the computations.

Define $\beta^T = (\beta_1, \tilde{\beta}^T)$, $\tilde{\gamma} = (\beta_1, \gamma_2, \ldots, \gamma_{K_n})$, and let $\tilde{\beta}_\gamma$ include $\tilde{\beta}_j$'s ($j = 2, \ldots, K_n$) with $\gamma_j = 1$. Consider the following MCMC algorithm starting from any initial position.

For $t = 1, 2, \ldots$:

(Step 1)  Sample $Z^t|\tilde{\beta}_\gamma^{t-1}, \tilde{\gamma}^{t-1}$.
(Step 2)  Sample $\tilde{\gamma}^t|Z^t$.
(Step 3)  Sample $\tilde{\beta}_\gamma^t|\tilde{\gamma}^t, Z^t$.

Below we explain each of the three steps and omit the time index $t$ to simplify notation.

*Step* 1. Note that $Z = (Z^{(1)}, \ldots, Z^{(n)})^T$. The step is carried out by independently sampling $Z^{(i)}$'s according to a "shifted" normal distribution:

1a: Generate $Z_i^* \sim N((x^{(i)})_\gamma^T \beta_\gamma, \sigma^2)$ independently where $v_\gamma$ denotes the subvector of $v_j$'s with $\gamma_j's$ being 1.

1b: Generate independent uniform variable $U_i^* \sim Unif[0, 1]$.

1c (Case 1): If $Z_i^* > 0$, set $Z^{(i)} = Z_i^*$ only when $U_i^* \leq a_+ = a_1/\max\{a_1, a_0\}$, where $a_{0,1} = p_{0,1}^{y^{(i)}}(1-p_{0,1})^{1-y^{(i)}}$.

1c (Case 2) : If $Z_i^* \leq 0$, set $Z^{(i)} = Z_i^*$ only when $U_i^* \leq a_- = a_0 / \max\{a_1, a_0\}$.

*Step* 2. Iteratively update one component at a time, conditional on all other components of $\tilde{\gamma}$. Define $Z^{(i)}(\beta_1) = Z^{(i)} - x_1^{(i)}\beta_1$, $Z(\beta_1) = (Z^{(1)}(\beta_1), \ldots, Z^{(n)}(\beta_1))^T$, $\tilde{X}_\gamma = (\tilde{x}_\gamma^{(1)}, \ldots, \tilde{x}_\gamma^{(n)})^T$.

2a: Simulate $\beta_1 | \gamma_2^{K_n}, Z$ to take value from $\pm 1$, with probability

$$p(\beta_1 | Z, \gamma_2^{K_n}) \propto 0.5 e^{0.5 \sigma^{-2} Z(\beta_1)^T [\tilde{X}_\gamma (\sigma^2 V_\gamma^{-1} + \tilde{X}_\gamma^T \tilde{X}_\gamma)^{-1} \tilde{X}_\gamma^T - I] Z(\beta_1)}.$$

2b: For $j = 2, \ldots, K_n$, simulate $\gamma_j | (\gamma_k)_{k \neq j}, \beta_1, Z$ to take value in $\{0, 1\}$, with probability

$$p(\gamma_j | \{\gamma_k : k = 2, \ldots, K_n, k \neq j\}, \beta_1, Z)$$
$$\propto \lambda^{\gamma_j} (1 - \lambda)^{1 - \gamma_j} I[|\gamma| \leq \bar{r}]$$
$$\times e^{0.5 \sigma^{-2} Z(\beta_1)^T [\tilde{X}_\gamma (\sigma^2 V_\gamma^{-1} + \tilde{X}_\gamma^T \tilde{X}_\gamma)^{-1} \tilde{X}_\gamma^T - I] Z(\beta_1)}$$
$$\times \{\det[I + \sigma^{-2} \tilde{X}_\gamma^T \tilde{X}_\gamma V_\gamma]\}^{-1/2}.$$

*Step* 3. Simulate $\tilde{\beta}_\gamma | \beta_1, \gamma, Z \sim N\{(\sigma^2 V_\gamma^{-1} + \tilde{X}_\gamma^T \tilde{X}_\gamma)^{-1} \tilde{X}_\gamma^T Z(\beta_1), \sigma^2 (\sigma^2 V_\gamma^{-1} + \tilde{X}_\gamma^T \tilde{X}_\gamma)^{-1}\}$.

*Note that all these conditional distributions are standard.* It can be easily shown that a stationary distribution of the proposed MCMC algorithm [which results in a Markov chain $(\gamma^t, \beta_\gamma^t)$ and its corresponding parameters $(\beta^t)$] is the desired Gibbs posterior $\omega(d\beta | D^n) \propto e^{-n\psi R_n} \pi(d\beta)$, where $R_n$ is the smoothed empirical risk in (RnR). We conjecture that the proposed MCMC algorithm converges to the desired Gibbs posterior in total variation distance, as $t \to \infty$, regardless of the starting position.

**8. Discussion.** The current paper studies a new Bayesian variable selection (BVS) method using a Gibbs posterior, which is directly constructed from a sample risk function of interest. This approach can perform better than the usual approach that uses a likelihood-based posterior, which in some situations can give a suboptimal risk performance with model misspecification. A smoothed sample risk function is used to provide convenient posterior computation in the style of Markov chain Monte Carlo. With BVS, the procedure can effectively handle high-dimensional data. We show that the resulting risk performance, even in a very high-dimensional case ($K \gg n$), can resemble the risk performance in a low-dimensional setting, in the sense that it can approach the best possible risk performance (achievable by certain sparse decision rules) at a low-dimensional convergence rate.

The approximately parametric/low-dimensional rate that BVS achieves, despite the high dimensionality ($K \gg n$), seems to defy the "curse of dimensionality." The reason is that BVS has used the so-called "bet-on-sparsity" principle [e.g., Friedman, Hastie, Rosset, Tibshirani and Zhu (2004)] by fitting effectively low-dimensional models due to the use of the prior distribution. Such a bet of course can be wrong: that is, we can be in the nonsparse case where all $x_j$'s can be important. However, in such cases not too much will be lost by the wrong bet, since nothing else seems to work well in such high-dimensional nonsparse case. On the other hand, when the bet is right, BVS can do much better than a minimax type rule that tries to protect for the bad cases. Intuitively speaking, a linear regression model without variable selection would have a large variance $K/n$ to start with, which is doomed to fail from the beginning, when $K \gg n$. On the other hand, BVS would use lower-dimensional submodels to make sure that the variance part is not out of control in the first place. When sparseness holds (i.e., only a few out of $K$ candidate $x_j$'s are important), the method will perform very well. It is noted that the sparse case can describe quite practical situations such as how a disease is mainly affected by only a few genes out of thousands.

A related approach to variable selection is based on Bayesian decision theory, which was described by Lindley (1968) and more recently extended by, for example, Brown, Fearn and Vannucci (1999), to the multivariate case. This approach is characterized by assuming normal data and using a friendly loss function (such as the quadratic loss). Under this framework, various expectations can be analytically computed and optimization can be simplified to depend only on the model indicator $\gamma$. Our approach cannot have such computational simplification and the Gibbs posterior needs to generate both $\gamma$ and $\beta_\gamma$ (parameter within the model). This is because we allow more general cases with a nonnormal predictor $x$ and nonquadratic loss. Our approach can handle classification error as well as realistic dollar-costs used in data mining. In addition, the current paper studies frequentist properties of risk performance which were not addressed in previous works using the Bayesian decision-theoretic approach.

The current approach generates a Gibbs posterior based on which both variable selection and model averaging can be performed. The theoretical result in the current paper is on a good performance of expected risk $E_{\beta,D} R(\beta) = E_D[E_{\beta|D} R(\beta)]$, which involves using models obtained randomly from the Gibbs posterior for $\beta|D$. [The parameter $\beta$ has certain nonzero components selected by a model indicator and determines a decision rule $I(x^T \beta > 0)$.] We argue that how to optimally utilize these good decision rules obtained from the Gibbs posterior (e.g., how model averaging can be done) is a nontrivial interesting problem. Model averaging would involve using rules parameterized by $E(\beta|D)$ instead of $\beta$. By Jensen's inequality, if $R(\beta)$ is convex, model averaging is always beneficial since $E_D[R(E(\beta|D))] \leq E_D[E_{\beta|D} R(\beta)]$. However, the classification error $R$ can be nonconvex and can have multiple minimums. In such cases the averaged decision rule can be a poor one even if each individual rule being averaged is good. It may be that some kind of model average "locally" can still be beneficial, in a limited region of approximate convexity, roughly speaking. This is worth further investigation.

The current approach uses a general framework allowing model misspecification (when the true generating process can be outside of the support of the prior). Although the proposed approach has an advantage in such a case with misspecification, we expect that in the case without misspecification (when the true model is within the support of the prior), the conventional approach using the likelihood-based posterior should perform comparably well to our procedure. This is because the conventional approach essentially minimizes the Kullback–Leibler (KL) divergence, which will lead to a good risk performance due to a relation between the two, when there is no misspecification. Such a relation is known, for example, between the classification risk and the KL divergence [see, e.g., Devroye, Györfi and Lugosi (1996), Problem 15.3].

Although this paper has focused on a smoothed sample risk for constructing the Gibbs posterior [choice (ii) in Section 2], similar results on good risk performance can be obtained when the unsmoothed sample risk [choice (i), or a sample version for the more general data mining risk described in Remark 1] is used. The proof will be more conventional and involve probability bounds for uniform deviation of sample risk based on the Vapnik–Chervonenkis theory [see, e.g., Devroye, Györfi and Lugosi (1996), Chapters 12 and 13 for a good description]. The posterior simulation can be based on the Metropolis algorithm [see, e.g., Tanner (1996), Chapter 6, for a description]. It is also noted that although we have focused on the Gibbs sampler in this paper, the Metropolis algorithm can also be applied to both cases with the unsmoothed and the smoothed sample risk. In the latter case with a smoothed sample risk, the Gibbs sampler approach of Section 7 may require a relatively large smoothing parameter ($\sigma$) for improving the algorithmic convergence. This would lead to some bias which can be corrected by applying the Metropolis algorithm with less (or no) smoothing.

**Acknowledgment.** We wish to thank an Associate Editor for the insightful comments and some additional references.

## REFERENCES

BROWN, P. J., FEARN, T. and VANNUCCI, M. (1999). The choice of variables in multivariate regres- sion: A non-conjugate Bayesian decision theory approach. *Biometrika* **86** 635–648. MR1723783

DEVROYE, L., GYÖRFI, L. and LUGOSI, G. (1996). *A Probabilistic Theory of Pattern Recognition.* Springer, New York. MR1383093

DOBRA, A., HANS, C., JONES, B., NEVINS, J. R., YAO, G. and WEST, M. (2004). Sparse graphical models for exploring gene expression data. *J. Multivariate Anal.* **90** 196–212. MR2064941

FRIEDMAN, J., HASTIE, T., ROSSET, S., TIBSHIRANI, R. and ZHU, J. (2004). Discussion on boosting. *Ann. Statist.* **32** 102–107.

GEMAN, S. and GEMAN, D. (1984). Stochastic relaxation, Gibbs distributions, and the Bayesian restoration of images. *IEEE Trans. Pattern Anal. Machine Intell.* **6** 721–741.

GEORGE, E. I. and MCCULLOCH, R. E. (1997). Approaches for Bayesian variable selection. *Statist. Sinica* **7** 339–373.

GERLACH, R., BIRD, R. and HALL, A. (2002). Bayesian variable selection in logistic regression: Predicting company earnings direction. *Aust. N. Z. J. Statist.* **44** 155–168. MR1963292

GREENSHTEIN, E. (2006). Best subset selection, persistency in high dimensional statistical learning and optimization under $\ell_1$ constraint. *Ann. Statist.* **34** 2367–2386. MR2291503

HOROWITZ, J. L. (1992). A smoothed maximum score estimator for the binary response model. *Econometrica* **60** 505–531. MR1162997

KLEIJN, B. J. K. and VAN DER VAART, A. W. (2006). Misspecification in infinite-dimensional Bayesian statistics. *Ann. Statist.* **34** 837–877. MR2283395

JIANG, W. (2007). Bayesian variable selection for high dimensional generalized linear models: Convergence rates of the fitted densities. *Ann. Statist.* **35** 1487–1511. MR2351094

LEE, K. E., SHA, N., DOUGHERTY, E. R., VANNUCCI, M. and MALLICK, B. K. (2003). Gene selection: A Bayesian variable selection approach. *Bioinformatics* **19** 90–97.

LINDLEY, D. V. (1968). The choice of variables in multiple regression (with discussion). *J. Roy. Statist. Assoc. Ser. B* **30** 31–66. MR0231492

SMITH, M. and KOHN, R. (1996). Nonparametric regression using Bayesian variable selection. *J. Econometrics* **75** 317–343.

TANNER, M. A. (1996). *Tools for Statistical Inference: Methods for the Exploration of Posterior Distributions and Likelihood Functions*, 3rd ed. Springer, New York. MR1396311

TANNER, M. A. and WONG, W. H. (1987). The calculation of posterior distributions by data augmentation (with discussion). *J. Amer. Statist. Assoc.* **82** 528–550. MR0898357

ZHANG, T. (1999). Theoretical analysis of a class of randomized regularization methods. In *COLT 99. Proceedings of the Twelfth Annual Conference on Computational Learning Theory* 156–163. ACM Press, New York. MR1811611

ZHANG, T. (2006a). From $\epsilon$-entropy to KL-entropy: Analysis of minimum information complexity density estimation. *Ann. Statist.* **34** 2180–2210. MR2291497

ZHANG, T. (2006b). Information theoretical upper and lower bounds for statistical estimation. *IEEE Trans. Inform. Theory* **52** 1307–1321. MR2241190

ZHOU, X., LIU, K.-Y. and WONG, S. T. C. (2004). Cancer classification and prediction using logistic regression with Bayesian gene selection. *J. Biomedical Informatics* **37** 249–259.

DEPARTMENT OF STATISTICS
NORTHWESTERN UNIVERSITY
EVANSTON, ILLINOIS 60208
USA
E-MAIL: wjiang@northwestern.edu
        mat132@northwestern.edu